# *PREPRINT[1]*:
# Coordinating Knowledge Work in Multi-Team Programs: Findings from a Large-Scale Agile Development Program

Torgeir Dingsøyr[1,2], Nils Brede Moe[1], Eva Amdahl Seim[3]

[1]*SINTEF Digital,* [2]*Department of Computer Science, Norwegian University of Science and Technology,* [3]*SINTEF Technology and Society*

---

# Abstract

Software development projects have undergone remarkable changes with the arrival of agile development methods. While intended for small, self-managing teams, these methods are increasingly used also for large development programs. A major challenge in programs is to coordinate the work of many teams, due to high uncertainty in tasks, a high degree of interdependence between tasks and because of the large number of people involved. This revelatory case study focuses on how knowledge work is coordinated in large-scale agile development programs by providing a rich description of the coordination practices used and how these practices change over time in a four year development program with 12 development teams. The main findings highlight the role of coordination modes based on feedback, the use of a number of mechanisms far beyond what is described in practitioner advice, and finally how coordination practices change over time. The findings are important to improve the outcome of large knowledge-based development programs by tailoring coordination practices to needs and ensuring adjustment over time.

# Introduction

Software development has undergone remarkable changes with the arrival of agile development methods in the late 90ies (Dingsøyr, Nerur, Balijepally, & Moe, 2012). The agile methods put emphasis on customer involvement, technical product quality, incorporating changing and emerging requirements and stating that software development is best done in small, co-located and self-managed teams (Hoda, Noble, & Marshall, 2012). These methods have led to far-reaching changes in how software projects are planned and managed.

One of the major changes is the increased focus on software development as teamwork (Melo, Cruzes, Kon, & Conradi, 2013; Moe, Dingsøyr, & Dybå, 2010), with emphasis on arenas for planning, synchronization and review and on practices to make teams work efficiently together such as establishing shared code ownership and discussing and learning through practices such as programming in pairs.

From being used for small co-located teams, the agile methods are increasingly used also in other settings such as in large programs with multiple teams (Xu, 2009). Large programs in general incorporate technical and organizational complexity. This includes a large number of stakeholders, a large number of program participants, a large number of requirements, lines of code and often very complex interdependencies between tasks as well as teams that depend on other teams. Programs using agile methods risk a lack of interaction and difficulties in communication (Xu, 2009). Such complexity has in general been found to have a negative effect on project performance (Floricel, Michela, & Piperca, 2016). Large-scale programs pose a great risk and are often associated with cost overruns, late completions, and project failures (Flyvbjerg, 2014; Flyvbjerg & Budzier, 2011).

The success of large programs is dependent on the program's ability to manage this complexity. As work is carried out simultaneously by many developers and development teams, coordination is of paramount importance (Fagan, 2004). Coordination is more important to team performance in large projects (Hoegl, Weinkauf, & Gemuenden, 2004) than in one-team projects. Therefore it is important to study how coordination practices are used in large-scale agile development. The literature on coordination has given much attention to permanent constellations such as organizations, but less to temporal constellations such at projects and programs (Dietrich, Kujala, & Artto, 2013).

This study describes agile practices in a large, multi team program, focusing on how the practices enable coordination on inter-team, project and program level. We describe the development method as a blend of agile and traditional methods, and discuss how this combination served to improve coordination. We address the following research questions:

1. *How are coordination practices used in large-scale agile development programs?*
2. *How do coordination practices change over time?*

The understanding of coordination in large programs is currently limited (Dietrich et al., 2013). Software development programs have developed new ways of working which could provide relevant insight for other types of knowledge-intensive projects (Conforto, Salum, Amaral, da Silva, & de Almeida, 2014; Serrador & Pinto, 2015). Also, agile development in large scale challenge assumptions in existing methods (Rolland, Fitzgerald, Dingsøyr, & Stol, 2016). Further, large programs are often critical for our society and today most advice on conducting such programs are based on experience rather than research. This study offers rich descriptions of use of concrete practices. These practices add to what is described in existing advice on agile software development. Finally, how coordination changes over time has been given little attention in research literature so far (Jarzabkowski, Le, & Feldman, 2012), understanding changes in practices will enable participants in programs to adjust coordination practices to needs. We position this research in line with thoughts on *rethinking project management*, focusing on handling the complexity of projects, and aiming at developing theory for practice (Winter, Smith, Morris, & Cicmil, 2006).

## Large-Scale Agile Development

Software development is a non-routine activity, as most systems are developed in one-of-akind projects. Software development is often described as creative work where a single optimal solution may not exist and progress towards completion can be difficult to estimate (Kraut & Streeter, 1995). One reason is that interdependencies between different tasks may be uncertain or challenging to identify, making it difficult to know who should be involved in work, and whether there is a correct order in which parties should complete their own specialized work (G. A. Okhuysen & B. A. Bechky, 2009). Changes in customer needs and in technology has also posed challenges for software development projects, and emphasized other needs for project management than what is found as engineering practices in other domains (Bryant, 2000).

Agile development methods is an umbrella term for a range of methods (Pekka Abrahamsson, Oza, & Siponen, 2010) sharing a set of key ideas formulated in the agile manifesto. We define agile methods as development methods that "rapidly or inherently create change, proactively or reactively embrace change, and learn from change while contributing to perceived customer value (economy, quality and simplicity)" (Conboy, 2009).

The most widely used agile method so far is Scrum (Rising & Janoff, 2000; Schwaber & Beedle, 2001), and this method is also the one that provides most advice on how to manage a development project (Pekka Abrahamsson et al., 2010). However, as a development team is self-managing, the project manager role is removed, and the only roles in the team is "developers" and a team facilitator, the "Scrum Master". The Scrum master is in charge of solving problems that prevents the Scrum team (5-9 people) from working effectively. The Scrum master works to remove the impediments of the process, runs and ensures decision making in the daily meetings and validates them with management (Schwaber & Beedle, 2001). Software is developed by the self-managing team in iterations called "Sprints", starting with a planning meeting and ending with a review with demonstration of the product and a retrospective focusing on process improvement. During a Sprint, a team coordinates through daily meetings, often in front of a "Scrum Board". Features to be implemented are registered in a product backlog, as "user stories" that should be understandable by the customer organization. User stories are often grouped into broader "epics". The "Product Owner" provides priority on backlog items in dialogue with the team. The tasks to be performed in the next iteration are listed in the "Sprint Backlog". Multiple stakeholders can participate in generating product backlog items, such as customer, project team, marketing and sales, management and support (P. Abrahamsson, Salo, Ronkainen, & Warsta, 2002).

Recently, there has been an increasing attention to how agile methods can be used in large development project or programs. We define "very large-scale agile development" as "agile development efforts with more than ten teams" (Dingsøyr, Fægri, & Itkonen, 2014), which have complex knowledge boundaries within the program. Further, such programs are characterized by a complex interplay with a larger number of technologies involved and usually a large set of stakeholders (Rolland et al., 2016).

There is a small body of studies on team coordination in very large-scale agile development, such as (Xu, 2009). Vlietland and Vliet (2015) propose that embedded coordination practices within and between Scrum teams positively impact delivery predictability in large projects. A study of "Scrum of Scrums" (Paasivaara, Lassenius, & Heikkila, 2012) suggests that this forum did not lead to satisfactory coordination: feature-specific or site-specific fora were better, but coordination at the project level was still a challenge. Researchers working closely with SAP (A. Scheerer, Hildenbrand, & Kude, 2014; Alexander Scheerer & Kude, 2014) have developed models of coordination called "coordination configurations" and are exploring how coordination configuration influences coordination effectiveness. Paasivaara and Lassenius (2014) describe a very large-scale development initiative at Ericsson with 40 teams where four types of communities of practice (Wenger, 1998) are used to coordinate teams. A survey on coordination in large-scale software teams found that respondents wished more effective and efficient communication, as well as the importance of good personal relationships for coordination (Begel, Nagappan, Poile, & Layman, 2009).

## Coordination

Coordination can be understood as "management of interdependencies between activities" (Malone & Crowston, 1994) and coordination mechanisms are the organizational arrangements, which allow individuals to realize a collective performance (G. A. Okhuysen & B. A. Bechky, 2009). Interdependencies include sharing of resources, synchronization of activities, and prerequisites activities.

Basic mechanisms for coordination in management science (Mintzberg, 1989) include: direct supervision, mutual adjustment, and standardization of work, outputs, skills and norms. Direct supervision is when one person is responsible for coordinating the work and give directives to those who are to do the work. Mutual adjustment is when workers adjust themselves to each other as their work proceeds. The other mechanisms are different kinds of pre-planned standardization: standardization of work, output, skills and knowledge, and norms.

Knowledge-intensive work like developing services based on software, brings a new sense of acuteness to the coordination challenge and the need of awareness because the speed of innovation change invalidates pre-determined interdependencies (Ramesh, Pries-Heje, & Baskerville, 2002). In such work, team members need mutual awareness to coordinate themselves by adjusting their own work to the work of others. Research has proposed different conceptual approaches for such adjustments, for example transactive memory systems (Wegner, 1986), sensemaking (Karl E. Weick, 1995), shared cognition (Cannon-Bowers & Salas, 2001), Complex Adaptive Systems (CAS) (Vidgen & Wang, 2009), collective problem solving (Hutchins, 1991; K. E. Weick, Sutcliffe, & Obstfeld, 1999) and collective mind (Crowston & Kammerer, 1998). These studies, and studies on expertise coordination (Faraj & Sproull, 2000) contribute to the insight of how team members can coordinate their actions in response to what other team members or people outside the team are doing.

Agile methods were designed to cope with change and uncertainty for small teams. These methods "de-emphasize traditional coordination mechanisms such as forward planning, extensive documentation, specific coordination roles, contracts, and strict adherence to a pre-defined specified process" (Strode, Huff, Hope, & Link, 2012) and mainly promote informal coordination (Xu, 2009). Agile development methods "embrace" change by moving decision authority to the team level, making rough long-term plans and detailed short-term plans. In their article entitled "why scrum works", Pries-Heje and Pries-Heje (2011) states that Scrum "requires very little time trying to foresee and negotiate the work flow and coordination mechanisms prior to actually conducting the work". They emphasize four artifacts that they believe are especially useful for coordination: The product backlog, the sprint backlog, the scrum board and the daily meetings. Strode et al. (2012) provides a comprehensive review of coordination studies in agile development, and have developed a model of coordination in agile software development projects (at team level), describing coordination strategies in terms of synchronization (activities and artifacts), structure (proximity of team members, availability of team member, substitutability of team members) and boundary spanning (interaction with other organizations outside of project). A particular mechanism to facilitate synchronization is the length of iterations. Shorter iterations will increase coordination but at the cost of more frequent planning and review meetings. Two-week iterations are common in small project teams.

## *Coordination Modes*

In large projects and programs, the work is often given to teams. Several factors then define the need for coordination between the teams. Van de Ven et al. (1976) discusses three main determinants of coordination mechanisms for organizations:

- *Task uncertainty* - the "difficulty" and "variability" of work undertaken by an organizational unit. Higher degrees of complexity, thinking time to solve problems or time required before an outcome is known indicates higher task uncertainty.
- *Task interdependence* - the extent to which persons in an organizational unit depend on others to perform their work. A high degree of task-related collaboration means high interdependence.
- *Size of work unit* - the number of people in a work unit. Increases in participants in a project or program means an increase in work size unit.

There are a number of mechanisms that can be applied to achieve coordination, and coordination is usually exercised through several mechanisms (Dietrich et al., 2013). Van de Ven et al. (1976) proposes three coordinating modes, which is used by Dietrich (2013) in their study of multi-team projects. The first two are based on feedback (or "mutual adjustment" (Mintzberg, 1989)), while the last is based on codification:

*Table 1: Coordination modes, definition and main coordination mechanisms* (Dietrich et al., 2013).

| Coordination mode | Definition (Dietrich et al. 2013) | Coordination mechanism (van de Ven et al. 1976) |
|---|---|---|
| Group mode of personal coordination | Use of mechanisms in which mutual adjustments occur in a group of occupants (more than two) through meetings | Scheduled meetings Unscheduled meetings |
| Individual mode of personal coordination | Use of mechanisms in which individual role occupants make mutual task adjustments through vertical or horizontal communication | Horizontal channels Vertical channels |
| Impersonal mode of coordination | Use of a codified blueprint of action that is impersonally specified | Blueprints of action |

*Group mode*, the mechanism for mutual adjustment (Mintzberg, 1989) is vested in a group of role occupants through scheduled or unscheduled meetings. Scheduled meetings are usually used for routine meetings, involving planned communication, while unscheduled meetings are used for unplanned communication between more than two participants. In agile development, group mode coordination at team level is ensured through sprint planning meetings, daily scrum meetings, sprint demonstration meetings and retrospectives (Strode et al., 2012; Xu, 2009).

*Personal mode*, where individual role occupants serve as the mechanism for making mutual task adjustments through either vertical or horizontal channels of communication. In horizontal channels, the "linkage function is assumed by an individual unit member who communicates directly with other role actors on a one-to-one basis in a non-hierarchical relationship" (Van de Ven et al., 1976). The mechanisms for vertical communication are usually line managers and unit supervisors. In large programs, this would be program management, project and sub-project managers and team leaders. In agile development, practices in extreme programming (Beck & Andres, 2004) such as pair programming, co-location, shared code ownership (Strode et al., 2012) and on-site customers (Xu, 2009) support horizontal coordination.

*Impersonal mode,* the coordination mechanisms are "programmed" or codified, and once implemented, they require minimal verbal communication between people. Examples are pre-established plans, process documentation, intranet pages, information technology tools and roadmaps. A "codified blueprint of action is impersonally specified" (Van de Ven et al., 1976, p. 323). This is present in agile methods such as in coding standards (Xu, 2009), but we can also see agile methods themselves as a type of impersonal mode, the method Scrum codifies types of meetings, roles and sets expectations for stakeholders.

Changes in coordination practices have been found to have significant influence on information sharing, work flow fluency between teams, efficiency of projects and learning outcomes (Dietrich et al., 2013).

As determinants change, prior studies indicate corresponding changes in coordination mode. Van de Ven et al. (1976) found that increase in task uncertainty leads to a substitution of the impersonal coordination with horizontal coordination mechanisms and group meetings. This gives a need for extensive and dynamic knowledge exchange to solve problems and adjust for emerging changes (Van de Ven et al., 1976). Dietrich (2013) also point to prior studies, which found that technological novelty relate to a higher rate of group meetings instituted by management. Project managers can achieve more control of work in such uncertain situations by relying on group-driven interaction in scheduled meetings.

Increased interdependence among persons in units in general leads to increased use of personal modes of coordination (Dietrich et al., 2013) and in particular the individual mode (Van de Ven et al., 1976).

Increased unit size, however, is associated with greater use of impersonal coordination and hierarchy (but no decrease in group mode coordination) (Dietrich et al., 2013).

In their study of multi-team projects, Dietrich et al. (2013) describe an information systems project in addition to five cases from other domains. The project had three concurrent teams as well as a project manager, steering group, a quality control group, a coordination group and also a one-person project office. Teams had a dedicated team leader. There is no information about development process in the case description. This project was characterized by a high degree of use of personal coordination modes, especially with a high use of the individual coordination modes. The study also reports use of some mechanisms in the impersonal mode.

In addition, large programs are temporal constructions where there is a large need to learn as everyone is new to the program. Typically in development programs, developers will need to learn about the business domain, and in addition there are constant developments in technology and work methods, which requires learning. It is therefore also interesting to investigate changes over time:

### *Change over Time: from Coordination to Coordinating*

Recent studies on coordination mechanisms have criticized prior studies for adopting a static view on coordination. A static or fixed view of coordination mechanisms adopted in prior research has limitations (Crowston, 1997; Jarzabkowski et al., 2012; Gerardo A Okhuysen & Beth A Bechky, 2009), focusing on activities that can be measured at a point in time. Jarzabkowski et al. (2012) refer to research, which shows that coordination mechanisms adjust to adapt to uncertainty, novelty and change over time. They argue that coordination mechanisms are "dynamic social practices that are under continuous construction" and describe how coordination mechanisms change over time. In uncertain situations with major changes, hierarchies and rule-based systems have been found to be less useful. In such situations, informal, interpersonal communications have been found to have larger influence on coordination.

## Method and Case

This study builds on a broader revelatory case study, which investigates how agile methods can be adapted in the very large scale. That article (Dingsøyr, Moe, Fægri, & Seim, 2017) focuses on how a large development program dealt with technical architecture, customer involvement and inter-team coordination. We have taken the material from that case and further analyzed our data material on coordination by using the framework established in Table 1 and the added dimension of change over time. We refer to the broad article for a complete description of data collection and analysis procedures, but provide arguments for case selection, and brief descriptions of data collection and analysis here. In addition, we provide an overview of the case.

The case was chosen because it was described by practitioners as a successful very large program that used agile development methods to a large degree. The whole program was co-located[2], and coordination mechanisms could be studied in a setting, which is well suited for agile methods. The "Omega" program developed a new office automation system for a public department. The program was managed by the department and involved two main consulting companies as subcontractors in the project *development*. The program ran from 2008 to 2012 and at the most, 12 teams were working in parallel on development, in total 175 people.

Our data collection started when the program was finished, focusing on group interviews and documents. We organized group interviews with the public department and the two consulting companies for each of the themes "architecture", "customer involvement" and "inter-team coordination and knowledge-sharing". 24 program participants took part in these 12 interviews, each interview lasting two hours, producing 247 pages of transcribed material. The documents were an official report after program completion, an internal experience report and the quality assurance report from the program. These reports contain 277 pages of text. The interview guide for inter-team coordination is provided in Appendix 1.

We analyzed the material in a tool for qualitative analysis, by using the framework established in the background section to identify expressions relating to coordination modes and changes in coordination mechanisms over time.

The main arguments for initiating Omega were public reform that required new functionality in office automation, and the existing office automation system was on a platform that was to be abandoned. When the program started, the content of the public reform was not known. This was the main reason for choosing agile development practices for the project. The public department has about 380 employees and provides 950,000 customers with several types of services. The department integrates heavily with one other public department.

---

[2] From about a year into the programme.

Omega was initiated to enable the department to provide accurate and timely services to customers and ensure a cost-effective implementation of the reform. Because of the reform, the program had a strict deadline. It is one of the largest IT programs in Norway, with a final budget of about EUR 140 million. The program started in January 2008 and lasted until March 2012. Of the 175 people involved, 100 were external consultants from five companies. The program used both time and material and target price contracts for subcontractors. About 800,000 person-hours were used to develop around 300 epics, with a total of about 2,500 user stories. These epics were divided into 12 releases.

The program was managed by a program director who mainly focused on external relations, a program manager focusing on the operations, and a controller and four project managers responsible for the *architecture, test, business* and *development* projects (see Figure 1):

- *Architecture* - Responsible for defining the overall architecture in the program and for detailing user stories in the solution description phase. Consisted of a lead architect and technical architects from the feature teams. Two main suppliers participated on a time & material basis.
- *Test* - Responsible for testing procedures and approving deliverables from the development teams. Consisted of a lead tester and test resources from development teams.
- *Business* - Responsible for analysis of needs through defining and prioritizing epics and user stories in a product backlog. This project was manned with product owners and a total of 30 employees[3] from the line organization in the department. Functional and technical architects from development teams also contributed to this project.
- *Development* - Divided into three subprojects: one led by the public department (6 teams) with their own people and people from five consulting companies, and the two other subprojects led by external consulting companies (3 teams) and (3 teams). The feature teams worked according to Scrum with three-week iterations, delivering on a common demonstration day. The feature teams had roles such as Scrum master, as listed in Table 2. In addition to the 12 feature teams, the project had an environment team responsible for development and test environments.

---

[3] The number of people involved in the project varied; we use numbers in the peak period in the project from 2009 to 2011.

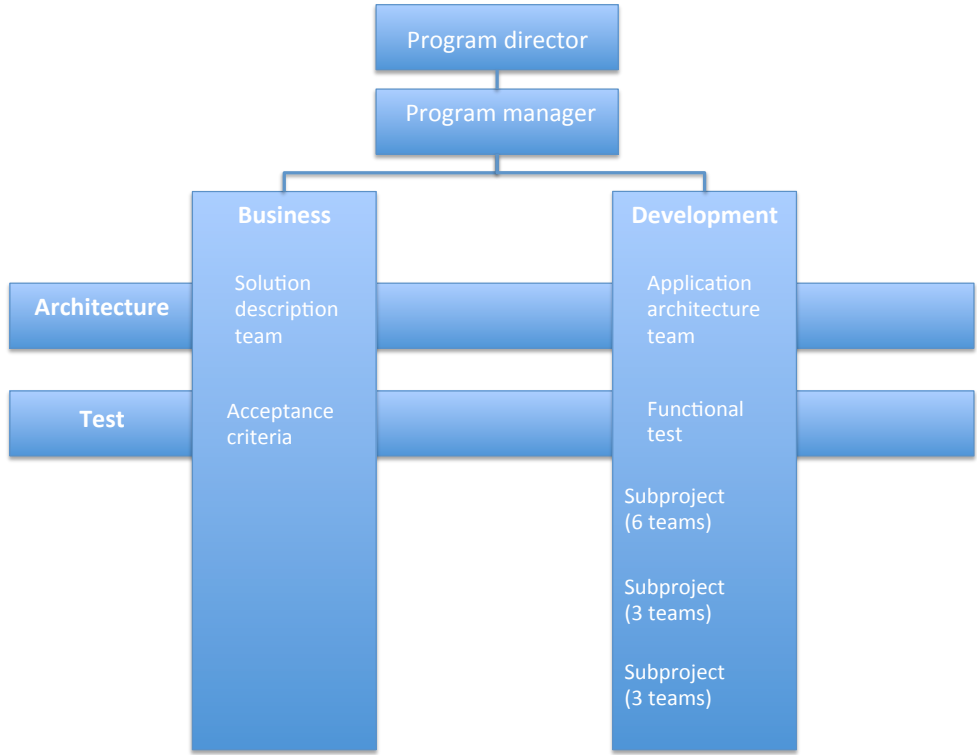

*Figure 1: The Omega Program with four main projects: Business, Architecture, Development, and Test. 12 development teams worked in the program at the peak. From 2009 to 2011 more than 150 people were working in the program. Some participants in project development also participated in project Business.*

*Table 2: Roles in the feature teams.*

| Role | Description |
|---|---|
| Scrum master | Facilitated daily meetings, iteration planning, demonstration and retrospective. |
| Technical architect | Responsible for technical design, working 50% on this and 50% on development. Participated in project architecture. |
| Functional architect | Responsible for detailing of needs. This role was usually allocated 50% to analysis and design, and 50% to development. Participated in project business. |
| Test responsible | Made sure that testing was conducted at team level: unit tests, integration tests, system tests and system integration tests. Delivered test criteria to the project test. |
| Developers | 4–5 developers were allocated to a team (a mixture of junior and senior developers). |

There were also projects for communication and adoption to prepare users for the new systems, in total six projects.

As shown in Figure 1, the program used a matrix structure where the business and development projects took part in the architecture and test projects. This matrix structure meant that a feature team would then mainly take part in project *development*, while also devoting resources to projects *architecture* (through the technical architect), *business* (through the functional architect), and *test* (through the test responsible) as shown in the roles in Table 2.

The program started working at the main office of the public department, but in 2009, it was moved to a separate office building located in the same city. Here, all teams were organized around tables.

Initially, the development process included four phases:

- *Analysis of needs* - Walkthrough of the target functionality of a release and identification of high-level user stories. Product owners prioritized the product backlog.
- *Solution description* - The user stories were assigned to epics, and the user stories were described in more detail, including design and architectural choices. User stories were estimated and assigned to a feature team.
- *Construction* - Development and delivery of functionally tested solutions from the product backlog. Five to seven three-week iterations per release. The teams used Scrum, with Sprint Planning, daily meetings, Sprint Demonstration and Sprint Retrospectives.
- *Approval* - A formal functional and non-functional test to verify that the whole release worked according to expectations. This included internal and external interfaces as well as interplay between systems.

To ensure development work on high priority user stories, there was pressure to have solution descriptions ready for the feature teams. This meant that releases were constantly being planned, constructed, and tested. Thus, given the roles in Table 2, a feature team would constantly be engaged in construction for the current release, approving delivered functionality in the previous release, and analyzing needs for the next release. After approval from the program, new releases were acceptance tested, set in production, and underwent an approval phase before being accepted by the operational IT section of the department.

# Results

We provide an overview of the main coordination modes used in the program, the group mode of personal coordination, the individual mode of personal coordination and the impersonal mode of coordination. Coordination mechanisms found in these modes are shown in Table 3. In addition, we describe how the coordination mechanisms changed over time:

*Table 3: Coordination mechanisms after coordination mode.*

| Coordination mode | Findings |
|---|---|
| Group mode of personal coordination | Board discussions |
| | Demo |
| | Experience forum |
| | Lunch seminars |
| | Metascrum |
| | Open space technology |
| | Retrospectives |
| | Scrum of Scrums |
| | Technical corner |
| Individual mode of personal coordination | Rotation of team members |
| | Customer on-site |
| | Direct communication in open work area |
| Impersonal mode of coordination | Instant messaging |
| | Masterplan |
| | Open work area |
| | Architectural guidelines |
| | Team routines |
| | Cross-team routines |
| | Solution descriptions in wiki |

### *Group Mode of Personal Coordination*

The program was characterized a number of scheduled meetings as well as arenas for unscheduled meetings for coordination in groups. We first describe scheduled meetings at program and project levels:

At program level, the only arena where everyone would meet was at the demonstration meetings, which were held every three weeks. In addition, the program management met two times a week in a forum, which was called "Metascrum". The Metascrum included managers from the main projects and the central program management, giving attention to "high-level" obstacles to progress and assessment of risks in the program. Well into the program, a new arena was introduced, the "open space technology". Open space was a way to get the whole program to set up a number of meetings across the project organization in order to discuss challenges and improvement initiatives. The decision to use a chatting tool was a result of these meetings. In addition, there were separate meetings to identify dependencies in tasks before work was assigned to teams.

At project level, there were three main types of scheduled meetings: The meetings prescribed by the agile method Scrum, meetings in the main projects in the program, and fora at project level to share experience across the development teams.

Scrum of Scrums were held in the three development subprojects with Scrum masters and subproject managers from 3-6 development teams. Project managers sometimes participated in these meetings. One subproject had daily Scrum of Scrum meetings in the beginning, but reduced the frequency to three times per week. A topic discussed here were resources, "now we have two people who are ill in the team, and we have given away a person to the environment team, how shall we manage to deliver our stories in the iteration?" (subproject manager). Also, retrospectives were sometimes held across teams in the subprojects, but overall this was an activity within each team.

The main projects architecture, business and test had meetings with their own staff and the people who held roles in the development teams. In the business project, much of the work concentrated on managing dependencies, "there were dependencies throughout the program" (technical architect). One of the participants in meetings in the business project said, "when we talked to the product owner, the product owner said, "we need you to do this", but then we had to explain that to achieve that we first need to do these tasks" (functional architect). The meetings in project architecture focused on establishing architectural guidelines, but also focused on coordinating work amongst the development teams to reduce the number of teams working on the same part of the codebase. "This was to reduce the possibility of making trouble for each other - which we did". The codebase was organized to reduce these challenges and in meetings teams declared that "this is our central area of work this period, so please limit work in that area" (technical architect).

Experience-sharing across teams were the focus of several scheduled meetings at sub-project level: "Experience forum", "Lunch seminars" and "Technical corner" are examples of meetings that existed during the program. A topic discussed at the experience forum, was how to liven up the retrospectives, this was then a topic discussed amongst all participants in the development teams in one project. Participation in these meetings was voluntary.

Unscheduled meetings were easy to organize due to the open workspace. Unplanned meetings frequently took place around the boards that were available for each team. These were used to "discuss solutions, draw and make sketches" (subproject manager). These discussions spanned development teams and roles. The project management was placed on tables so that they could see most of the boards and thus quickly get an overview of status of the teams. If the project managers noticed discussions, they could inquire about the issue and say that "this problem I know was addressed by another team two iterations ago, let us get Ola over here and see if he can help" (subproject manager). A Scrum master and developer stated that they learned "very much" in the program during these discussions around the boards, but it was important to have sufficient coordination arenas so that people realize that "we need to talk". The program also started to use a group chatting tool (Jabber) in order to ease informal coordination, what we can see as a type of unscheduled virtual meetings. This tool was introduced during the program, which enabled asking several people for help without interrupting them. This channel was used for a number of purposes, from asking technical questions to informing about the next wine lottery.

Informants emphasized the importance of the unscheduled meetings. One said, "I think the combination of scheduled and unscheduled coordination that just appeared was very important" (scrum master and developer).

## *Individual Mode of Personal Coordination*

The program was characterized by direct informal coordination between members of different teams, using both horizontal and vertical channels:

The development program used a number of different arenas to coordinate work and share knowledge between teams. During the build-up phase, enrolment of new team members was frequent and of great importance. The program facilitated this by occasional splitting of existing teams and even distribution of new team members. Change of team members were also done to alleviate problems of personal chemistry. The frequency of both types of changes to teams was considerably lower in later phases. Team changes were an important and consciously used instrument for distributing knowledge and facilitating coordination, both horizontal and vertical. Over time, resistance to team changes increased markedly with strengthening team feeling: "there was a limited number of people who were candidates for such change due to competence, so I had to do some pep talks and get people to think positively". In order to facilitate self-organized teams, the development program sought to limit the authority of scrum masters compared to normal practice. A key motivation was to inspire the team members to take responsibility and coordinate internally and between teams. Scrum-masters were, however, key to effective changes to team composition, and to gather information through talking to all the team members to know the status of the work.

Horizontal coordination between team members changed over time. Many emphasized the importance of informal coordination, facilitated by the open work area. Team-members asked for advice across the team and organizations; "we are here to succeed and no one can succeed alone". Team members experienced personal coordination as crucial for solving interdependencies between tasks and keeping the sprint schedule; which could been described as a direct contact between experts. As the program progressed, pragmatism in the allocation of tasks between suppliers became extensive. Eventually, one could just ask "can you help me with this", and receive an "ok. We'll help you with this now if you help us with something else in the future". The management also sought to rotate the team-members in a way that some of the team-members from the development team also participate in the solution description. In addition, extensive personal coordination was also possible because of all contractors were working towards the same goals. Social arenas such as common lunches, coffee breaks and other social happenings were important coordination mechanisms during the project.

One of the mechanisms for vertical coordination was "management by walking around". It was used to get status from the team, help the teams and to spread important information, e.g. if one team had solved a particular problem. A manager explained what could initiate a short trip over to a team: "When you see that the team has been sketching on the board for two hours, and then it is time for you to get up and check out what is going on". Often they were struggling deciding between different alternative solutions, and then you could help them taking a decision by providing information or viewpoints they did not know about. Another explained: "during the teams stand-up I sometimes stopped by and listened to what they were talking about."

A culture developed where decisions were discussed informally between relevant stakeholders and subsequently formalized.

*Impersonal Mode of Coordination*

The main impersonal coordination mechanisms were the main plan for the program, guidelines and checklist. They all changed regarding content and technology during the program.

The general plan included all work to be done described as epics. All epics and tasks were initially documented in an electronic spreadsheet. However, this spreadsheet was replaced because of two problems related to coordination: First, because of the size of the program it was difficult to get a good overview of the whole plan by using spreadsheet technology. Second, it was difficult to know what the latest version of the spreadsheet was because multiple versions were created and distributed using various channels.

About a year into the program, an issue tracker (Jira) replaced the spreadsheet. This new tool for coordinating was introduced together with a major revision of the plan. The new plan included 300 epics and 22 work packages. The 300 epics were later decomposed to 2.500 user stories with subtasks. Every team could follow the program progress in the tool. While the issue tracker was mandatory, used by all teams and regularly updated, all teams duplicated their tasks on stickers on a board close to the table where they were located. Each team had their own board with an overview of tasks the team was committed to solve during the next iteration. A task was written on a sticker and moved around when the status of the task changed.

While the issue tracker was essential for coordination of tasks on the program and project level, the physical board was important for coordination on the team level. Also management could easily see the status of the work going on in a team by just looking at the board. As one said "it takes two seconds to get an overview of status [in a team], and from my location [in the open work area] I could see almost all the boards, and then I would know what had happened at the end of yesterday [in each team]". Another explained: "it was seen as an important ceremony to move one sticker one the board. Changing the status in the issue tracker does not bring an applause" (subproject manager).

The issue tracker was used together with a tool for facilitating code reviews for coordinating work. When the review work was registered in the tool, there was a minimal need for verbal communication among the users.

All the process descriptions documents, guidelines, and checklist were documented in a wiki (a website that provides collaborative modification of its content and structure directly from the web browser). The wiki was available for everyone and mandatory to use. Examples of routines were team routines and routines describing cross team collaboration like the daily Scrum meeting and Scrum of Scrum meetings. Examples of guidelines for designers and programmers in the wiki were guidelines for graphical user interface design, how to use the programming language Java and how to perform specific programming tasks. The guidelines included tips and experiences written by others in the program. The tips and experience connected to a guideline were regularly updated. An outcome from a sprint retrospective could for example initiate a change in a guideline.

While most guidelines and checklists were defined before they were used, many were created on request. One example was a team that saw a need for new architectural guidelines during an iteration, which initiated the architects coming together to establish the new guideline so that the next team could use them. As one architect said: "it is better do define guidelines when someone needs them, instead of us trying do identify all needed guidelines up front".

In the post project review the use of guidelines and plans were evaluated. It was found that some were defined too late causing problems with e.g. the error handling. Not everyone followed the guidelines because they felt it made them inflexible. Another explanation for lack of use was related to the number of guidelines, rules and processes. The size of the program made it hard to get a full overview, especially for newcomers.

### *Change over Time*

On a team level, the main mechanisms for coordination remained constant during the program. However, on inter-team level there were a number of changes over time:

For inter-team coordination, several meetings and forums were established. The main benefit of these forums and meetings was to build knowledge of who knows what. When people started to get an overview of who to talk to, informants state they did not need the meetings anymore and started approaching people directly, or they arranged unscheduled meetings, or discussed by the coffee machine or by the boards. One said: "we stopped doing some meetings because we could replace them with shorter meetings or because we got to know each other, then we could just talk to each other".

## Discussion

We structure the discussion of our findings after our two research questions. First, we ask *how are coordination practices used in large-scale agile development programs?*

From our results we see that all three modes of coordination are used in the Omega program. The program is characterized by high uncertainty regarding the tasks, a high degree of task interdependencies and finally by a large unit size. Prior studies then suggest that this situation would call for more coordination, and indeed we identified a number of coordination mechanisms in use, across all three modes of coordination. Our study did not measure extent of use, we are only able to state that certain mechanisms were used in the program.

An increase in task uncertainty has been found to lead to a substitution of impersonal coordination with horizontal coordination mechanisms and group meetings (Van de Ven et al., 1976). Within-team horizontal coordination has been identified as a characteristics in agile projects (Xu, 2009). In Omega, we found a high presence of horizontal coordination also across teams as well as a number of scheduled meetings. The high interdependence among persons has been found to lead to an increase in personal modes of coordination. We identified many mechanisms that were widely used within these modes.

An interesting finding in our material is the gradual transition to unscheduled meetings in the group mode, and that the scheduled meetings were seen by informants as a prerequisite for this transition. It is likely that a high number of scheduled meetings early in the program established relations and knowledge of other people´s skills. Further, the matrix organization of the program with team members taking part in all four major projects involved a number of scheduled meetings with subsequent development of relations and knowledge. The combination of arenas prescribed in agile development such as the Scrum of Scrums, demonstrations and retrospectives gave room to bottom-up coordination, while the scheduled meetings in the projects architecture, business and test gave management control.

Finally, the unit size is associated with greater use of the impersonal mode and hierarchy. We found many impersonal mechanisms in use, but most informants focused on the group mode when describing coordination practices. However, the organization of the program with separate projects for architecture, business and test gave emphasis to establishing guidelines and rules across the program. Plans were made visible both a team and program level through showing status of tasks both on boards for the teams and in the issue tracker for aggregations and overall status.

In contrast to Dietrich et al. (2013) most of the mechanisms identified in our study relate to the group mode, while Dietrich et al. found most mechanisms to relate to the individual mode of personal coordination. A reason for this could be the focus on practices for coordination in our data collection, as we did not use a targeted data collection scheme for all three coordination modes.

Comparing our findings to prior work on coordination in agile development, we see that all four artifacts emphasized for coordination by Pries Heje and Pries Heje (2013) was used in Omega: The product backlog, the sprint backlog, the scrum board and the daily meetings. Following the model by Strode et al. (2011), we see that synchronization was ensured through a number of practices, such as setting the iteration length to three weeks and following the Scrum method on team level. The open work area and full-time engagement of program members contributed to structure, and the matrix organization provided program-internal boundary spanners. If we compare the work in our program to work in a single agile team, we find a number of additional traditional practices focusing both on forward planning through the business and architecture projects, as well as on documentation as represented by the test project and criteria for accepting a developed user story. We also find a number of additional roles on different levels, such as the functional and technical architects at team level and also project managers and other administrative roles at project and sub-project level.

Secondly, we ask *how do coordination practices change over time?*

Our findings are in line with the view that coordination mechanisms are not static, but dynamic structures that change over time. While our explorative material does not allow us to show detailed traces of changes over time, our material shows a number of scheduled meetings that existed for a while and then disappeared, such as the experience forum and the technical corner. Informants state that informal communication in the open work area increased over time as people got to know each other. Also, new mechanisms such as the open space technology and the instant messaging appeared. There were changes in rules and plans, from making use of traditional spreadsheets and documents in the initial phase of the program to establishing a new "Masterplan" in an issue tracker with details and rules of work procedures described in a wiki. We can speculate that there were two main drivers of changes over time: First, the domain of the program was unknown to most external consultants working on development, which required much learning about the domain itself, and about who in the customer organization that could answer questions. Second, the program scaled up, splitting teams into two. This was done in order to meet the strict deadline of the program, and also led to a renewed focus on learning later in the program.

Also regarding the use of agile methods, there were changes over time. The frequency of Scrum of Scrum-meetings changed during the program, in one subproject from daily to three times a week. Also, the retrospectives were mainly conducted at team level, but sometimes also at subproject level. Informants state that as most decisions were discussed informally towards the end of the program, these decisions were recorded in the daily meetings, the Scrum of Scrums or in the Metascrum.

This revelatory case study has several limitations: First, we have not been able to follow the program over time, but have conducted our data collection after the program was finished. Second, as an explorative study in a new area, our data collection as been broad, and we have for example asked about coordination practices but not explicitly about impersonal modes of coordination. Thus, our material on coordination mechanisms probably does not provide a complete overview of the mechanisms used in the program.

## Conclusion

We have described how coordination mechanisms are used in a large-scale agile development program, and how these mechanisms change over time. Our case program was characterized by high task uncertainty, a high degree of interdependence for tasks, and by a large number of people.

Our research develops three main insights, which we think is relevant to the project management community when adopting practices from agile development:

First, an increase in task uncertainty has been found to lead to a substitution of impersonal coordination with horizontal coordination mechanisms and group meetings. We have established a high presence of personal communication; both in the group mode and in the individual mode. Informants emphasized the importance of the open work landscape for horizontal personal coordination and this also made vertical personal coordination easier as project managers quickly could be informed of status of teams when involving in one-to-one discussions. Also, establishing a mixture of agile and traditional scheduled meetings was important to build knowledge and relations early in the program. There was a high number of scheduled meetings initially, but a gradual transition to unscheduled meetings. The meetings related to the agile method Scrum was kept throughout the program, and the iteration length remained at three weeks. The frequency of scheduled meetings is very important when balancing risk of developing unwanted functionality and costs of "ceremony" in the form of time spent on planning and review. Our study supports the finding that personal coordination is central to achieving inter-team coordination in large programs.

Second, there were a number of coordination mechanisms in use, spanning all three modes of coordination. Table 3 lists 19 mechanisms identified in the program. Traditional "large-scale" agile development in contrast, only explicitly focuses on the "Scrum of Scrums" as a mechanism for inter-team coordination. One mechanism was also duplicated: The plan existed on program level in an issue tracker, while each team kept a version of the plan on their board. This duplication helped to serve needs for plans at different level. This suggests that one coordination mechanism is not enough; efficient coordination depends on a variety of mechanisms.

Third, there were frequent changes in how coordination took place. Scheduled meetings were extensively used in the introductory phase of the program, but were at later stages replaced by unscheduled meetings. New mechanisms were taken into use such as needs changed during the program execution. Coordination practices change over time.

Future work should seek to develop further understanding regarding coordination modes and mechanisms in large development programs, particularly interesting topics would be to investigate how coordination mechanisms are tailored to the specific context of a program, as well as to further understand how coordination needs change over time.

We believe the main implications for practice from this study is to highlight the number of mechanisms in use in a successful program, and giving rich descriptions of such mechanisms, which provides a number of suggestions in addition to what is described in the agile development literature. Second, we would like to emphasize the role the open co-working space had in this case, as an efficient enabler of coordination. Lastly, we would like to underline the change in coordination needs over time, which suggest practices to reflect on and change the development method as a program progresses, such as in the common practice in agile development of conducting retrospectives.

# Appendix 1: Interview guide

- How was the work organized in your part of the program?
- What kinds of dependencies were there between the teams in your part of the project? (examples?)
- How were dependencies managed? (examples?)
- What was managed in established fora and what was managed outside of the fora? (examples?)
- Who were involved in managing dependencies between teams? (examples?)
- Did you encounter challenges with managing dependencies? (examples?)
- Did you change the way you managed dependencies during the project? (examples?)
- What practices do you think were most important in order to manage dependencies between teams? (examples?)
- Are there any practices you think had little importance for managing dependencies?
- How did the division of the project into three main parts influence the coordination between teams?
- Were there differences in inter-team coordination across the subprojects?
- Frequency of meetings/persons involved/time spent